\documentclass[sigconf, authorversion]{acmart}
\usepackage{balance}
\usepackage[]{todonotes}
\def\BibTeX{{\rm B\kern-.05em{\sc i\kern-.025em b}\kern-.08emT\kern-.1667em\lower.7ex\hbox{E}\kern-.125emX}}
\usepackage{color}
\usepackage{booktabs}

\copyrightyear{2021}
\acmYear{2021}
\setcopyright{rightsretained}
\acmConference[ICPE '21]{Proceedings of the 2021 ACM/SPEC
International Conference on Performance Engineering}{April 19--23,
2021}{Virtual Event, France}
\acmBooktitle{Proceedings of the 2021 ACM/SPEC International
Conference on Performance Engineering (ICPE '21), April 19--23, 2021,
Virtual Event, France}
\acmDOI{10.1145/3427921.3450234}
\acmISBN{978-1-4503-8194-9/21/04}

\begin{document}


\title[Virtuous Cycle in Performance Testing]{Creating a Virtuous Cycle in Performance Testing at
  MongoDB}
\author{David Daly}
\orcid{0000-0001-9678-3721}
\affiliation{MongoDB Inc}
\email{david.daly@mongodb.com}

\begin{abstract}
  It is important to detect changes in software performance during development in order to avoid
  performance decreasing release to release or dealing with costly delays at release
  time. Performance testing is part of the development process at MongoDB, and integrated into our
  continuous integration system. We describe a set of changes to that performance testing
  environment designed to improve testing effectiveness. These changes help improve coverage,
  provide faster and more accurate signaling for performance changes, and help us better understand
  the state of performance. In addition to each component performing better, we believe that we have
  created and exploited a virtuous cycle: performance test improvements drive impact, which drives
  more use, which drives further impact and investment in improvements. Overall, MongoDB is getting
  faster and we avoid shipping major performance regressions to our customers because of this
  infrastructure.
\end{abstract}

\maketitle


\section{Introduction}\label{sec:intro}
Over the last several years we have focused on improving our performance testing infrastructure at
MongoDB\@. The performance testing infrastructure is a key component in ensuring the overall quality
of the software we develop, run, and support. It allows us to detect changes in performance as we
develop the software, enabling prompt isolation and resolution of regressions and bugs. It keeps
performance regressions from being included in the software we release to customers. It also allows
us to recognize, confirm, and lock in performance improvements. As a business, performance testing
impacts our top and bottom lines: the more performant the server, the more our customers will use
our services; the more effective our performance testing infrastructure, the more productive are our
developers. Testing performance and detecting performance changes is a hard problem in practice, as
performance tests and test platforms inherently contain some degree of noise. The use of change
point detection~\cite{daly_use_2020} was a large improvement in our ability to detect performance
changes in the presence of noise.

After putting our change point detection system into production, we explicitly focused on 4
challenges: how to deal with the large number of results and process all the changes; how to better
deal with and isolate noise due to the testbed system itself; how to easily compare the results from
arbitrary test runs; and how to capture and how to more flexibly handle more result types. The first
two are familiar challenges, having been an explicit focus of the change point detection work, while
the second two challenges become more serious problems once we achieved a basic ability to process
our existing results.

The cumulative impact of these changes and our previous work has been to enable a virtuous cycle for
performance at MongoDB\@. As the system is used more, we catch and address more performance changes,
leading to us using the system more.

The rest of this paper is organized as follows. In Section~\ref{sec:review} we review our previous
work on which this paper builds. In Section~\ref{sec:organic} we discuss changes that have happened
naturally as we have used the system more, leading to more load on the system. We then dive into
four changes that we have tried in order to improve our infrastructure:
Section~\ref{sec:betterprocess} for improving our processing of results,
Section~\ref{sec:descriptive} for handling more result types, Section~\ref{sec:systemissues} to
address system noise, and Section~\ref{sec:comparison} to improve the comparison of arbitrary test
runs. Those sections are followed by a dive into the practical impact of all these changes in
Section~\ref{sec:impact}, before reviewing future work, related work, and conclusions in
Sections~\ref{sec:future},~\ref{sec:related}, and~\ref{sec:conclusion}.

\section{Review}\label{sec:review}

We built our performance testing infrastructure to be completely automated, and integrated with our
continuous integration system Evergreen~\cite{noauthor_evergreen_nodate-1}. From past experience we
had concluded that it was essential to automate the execution and analysis of our performance tests,
and regularly run those tests as our developers worked on the next release. Previously we had done
ad-hoc testing and manual testing at the end of the release cycle. In both cases we were continually
challenged by test results that would not reproduce, as well as a huge diagnosis effort to identify
which component and changes to that component caused the performance changes. The combination of
those challenges led to a large effort late in each release cycle to try to identify and fix
performance regressions, often resulting in release delays or performance regressions shipping to
customers. Creating the infrastructure~\cite{ingo_automated_2020} to test performance in our CI
system let us identify and address regressions earlier, and made it much easier to isolate
performance changes.

Automation does not inherently make the tests reproducible, but it does make it clearer that there
is noise in the results. Further work went into lowering the noise in the test
results~\cite{henrik_ingo_reducing_2019}. That work lowered, but did not eliminate the noise in
the performance results. It was still challenging to detect changes in
performance. Originally we tested for performance changes above some threshold (usually $10\%$), but
this had a number of problems, leading us to use change point detection~\cite{daly_use_2020}. Change
point detection attempts to determine when there are statistical changes in a time-series, which is
precisely the problem we want to solve. After the transition to change point detection, we had a
system with completely automated, low noise tests that we could successfully triage and process.

\section{Organic Changes}\label{sec:organic}

There are a number of organic changes to our performance test environment that have occurred over
the last couple of years. These changes were not planned, but they were still important changes. The
performance testing system works, detecting that the performance has changed and correctly
identifying when those performance changes occurred. The development engineers have seen that it
works and so they use the performance test infrastructure more. One key aspect of that increase in
use is that the development engineers have added more tests. We have also added new test
configurations to further increase test coverage. Development engineers and performance
  engineers both add performance tests and configurations.

Table~\ref{tab:tests} shows the number of system under test configurations, tasks (collections of
tests), tests, and number of raw results from running the performance tests for any version of the
software. The data covers that past three years and is collected from tests run in September of each
year. The table specifically filters out canary\footnote{Canary tests are discussed in
  Section~\ref{sec:systemissues}.} results and anything that we would not actively triage. In some
cases, the line between configurations, tasks, tests, and results may be arbitrary, but it is how
our system is organized and users interact with each of those levels.

\begin{table}
  \centering
  \begin{tabular}{lrrr}
    \toprule
    & 2018 & 2019 & 2020 \\
    \cmidrule{2-4}
    Number of Configurations & 8 & 17 & 24 \\
    Number of Tasks & 86 & 181 & 356 \\
    Number of Tests & 960 & 1849 & 3122 \\
    Number of Results & 2393 & 3865 & 5787 \\
    \bottomrule
  \end{tabular}
  \caption{The number of total possible test results we can create per source code revision has
    increased significantly over the past two years. This is due to increases in the number of tests
    and the number of configurations in which we run those tests.}
  \label{tab:tests}
\end{table}

You can see the huge increase in every dimension. We run our change point detection algorithm on the time-series for
every result, and someone must triage all of that data. The total number of results went up
$50\%$ year over year, and $142\%$ over two years.

Additionally, the development organization has grown leading to more commits to our source
repository. Overall the number of engineers working on our core server has gone up approximately
$30\%$ year over year for the past two years. Table~\ref{tab:commits} shows the number of commits
and commits per day over the last 3 years. There has a been a steady increase in commits, going up
$18\%$ in the past year and $27\%$ over the past two years. Each commit can potentially influence
performance. If you combine the increased commit velocity with the increase in results per revision,
you get a $76\%$ increase in total results year over year, and an over $3x$ increase in total possible
results to generate and analyze over two years.

\begin{table}
  \centering
  \begin{tabular}{lrrr}
    \toprule
    12 months ending & 2018-09-01 & 2019-09-01 & 2020-09-01 \\
    \midrule
    Commits & 4394 & 4702 & 5538 \\
    Commits per day & 12.0 & 12.9 & 15.2 \\
    \bottomrule
  \end{tabular}
  \caption{The number of commits per day to our source repository has been increasing as the
    development organization has grown.}
  \label{tab:commits}
\end{table}

The net result of these changes (more commits + engineers using the system more) is many more
possible results that may introduce performance changes and need to be isolated. During this time we
have not increased the people dedicated to processing these results. All the problems we needed to
address in the past are increased. Our processes to find and isolate changes need to scale or they
will break down under the weight of new results.

\section{Better Processing of performance changes}
\label{sec:betterprocess}


In our previous paper~\cite{daly_use_2020} we described the role of ``build baron'': the ``build
baron'' is a dedicated role to triage all performance changes, producing JIRA tickets and assigning
them to the appropriate teams to address the changes. Originally the build baron role rotated
through the members of the team that built the performance infrastructure. On the positive side,
these people knew the system very well. However, that was balanced by the feeling that the work was
a distraction from their primary work. Build baroning was a large transition from regular day to day
work, and required both rebuilding mental state when becoming build baron and when returning to
normal work. Everyone tried to dedicate the proper time to the work, but it is easy to want to do a
little bit more of the development work you had been doing. Additionally, it's likely that the
skills for a build baron differ from the skills of a software developer.

As such, we built a new team dedicated to build baroning. This new team originally covered
correctness build failures, but has since expanded to the performance tests as well. The roles still
rotate with the build baron team, but the team is always doing triage (not triage and
development). The team members are better able to build up intuition and mental state about the
system, and can more easily get help from each other. Possibly more importantly for members of this
new team, triaging failures \textbf{is} their job, not an interruption from their job. While we
added this new team, we did not allocate more people to doing the build baroning, rather we shifted
who was doing the work.

The dedicated team is also able to better articulate the challenges of build baroning, and what
changes would make them more productive. Over time the team developed a set of heuristics to deal
with all the change points they had to process and shared knowledge. Part of this was adding filters
to the existing boards and new ways of looking at the data. Where feasible we reviewed these
heuristics and integrated them into the displays by default. Examples include better filtering of canary
workloads (recall we do not want to triage changes in canaries, but rather rerun them) and sorting
capabilities.

The impact of these changes show up in our overall statistics which are discussed in
Section~\ref{sec:impact}. The summary is that they allowed us to evaluate more tests and commits to
find more changes, while also increasing the overall quality of the generated tickets without any
additional human time.

\section{Making the System More Descriptive}\label{sec:descriptive}

Our performance testing environment was originally designed for tests that measured throughput, as
throughput based tests are the easiest to create and analyze (just run an operation in a loop for a
period of time, possibly with multiple threads). This assumption got built into the system. We knew
it was a limitation in our system and have been striving to get around it. We developed conventions
to add some latency results to our system, but it was inelegant. Worse, it largely assumed only one
result per test. Ideally we could measure many performance results per test, such as throughput,
median latency, tail latencies, and resource utilizations. Before change point detection, we could
not add significantly more metrics since we could not keep up with the simpler tests we already
had. Now that we had change point detection, we wanted to be able to track and process these
additional metrics.

There were fundamentally two ways we could add these new metrics: 1. Have tests measure the metrics
of interest and then compute and report the relevant statistics to the results system.  2. Have tests
measure the metrics of interest and report all of those results to the result system. In the second
case the test would report the metric for every operation --- much more data --- and let the results
system calculate the statistics. After some review, we decided we preferred case 2, but that we also
had to support case 1.

We preferred the more data intensive case 2 because of what it enables. If we run a test that
executes $10$k operations, the system will report the latency for each of those $10$k
operations. First, having all the data allows us to change and recompute the statistics in the
future. For example, if we decide we need the $99.99\%$ latency in addition to the existing
statistics, we can add it and recompute. If the test itself was computing the statistics we would
have to rerun the test. Additionally, it allows us to view performance over test time, within a test
and from the test's perspective (client side). This gives us a much more dynamic view of the
performance of the system. We chose our preferred case, and it was paired with work on our
open-source performance workload generation tool Genny~\cite{noauthor_genny_2020}. We created a new
service called Cedar~\cite{noauthor_package_nodate-1} to store the results and calculate the
statistics, and a tool called Poplar~\cite{noauthor_package_nodate} to help report the results from
the testbed to Cedar. Both are open source and part of our continuous integration system
ecosystem~\cite{noauthor_evergreen_nodate}.

While we chose the detailed case, we decided we also had to support the case in which tests computed
their own statistics. The reason for this was simple: in addition to workloads written in Genny, we
also run third party industry standard benchmarks in our regression environment (e.g.,
YCSB~\cite{cooper_benchmarking_2010, noauthor_ycsb_2020}). Those tests already generate their own
statistics, and it is not reasonable to adapt each such workload to report the metrics for every
operation. The system we built handles both the case of getting all the raw results and the case of
receiving the pre-aggregated data.

The new system was just that, a new system. We needed to integrate it into our production systems
without breaking anything. The test result history is important both to the result display as well
as the change point analysis, so we could not just turn off the old system and turn on the
new. Instead we needed to make the old system and the new work together in the UI\@. We also needed
to make it possible to handle the increase in information without completely overwhelming the build
baron team\footnote{The results discussed in this section are in addition to the increase in results
  discussed in Section~\ref{sec:organic}}.

\begin{figure*}[tb]
  \centering
  \includegraphics[width=\linewidth]{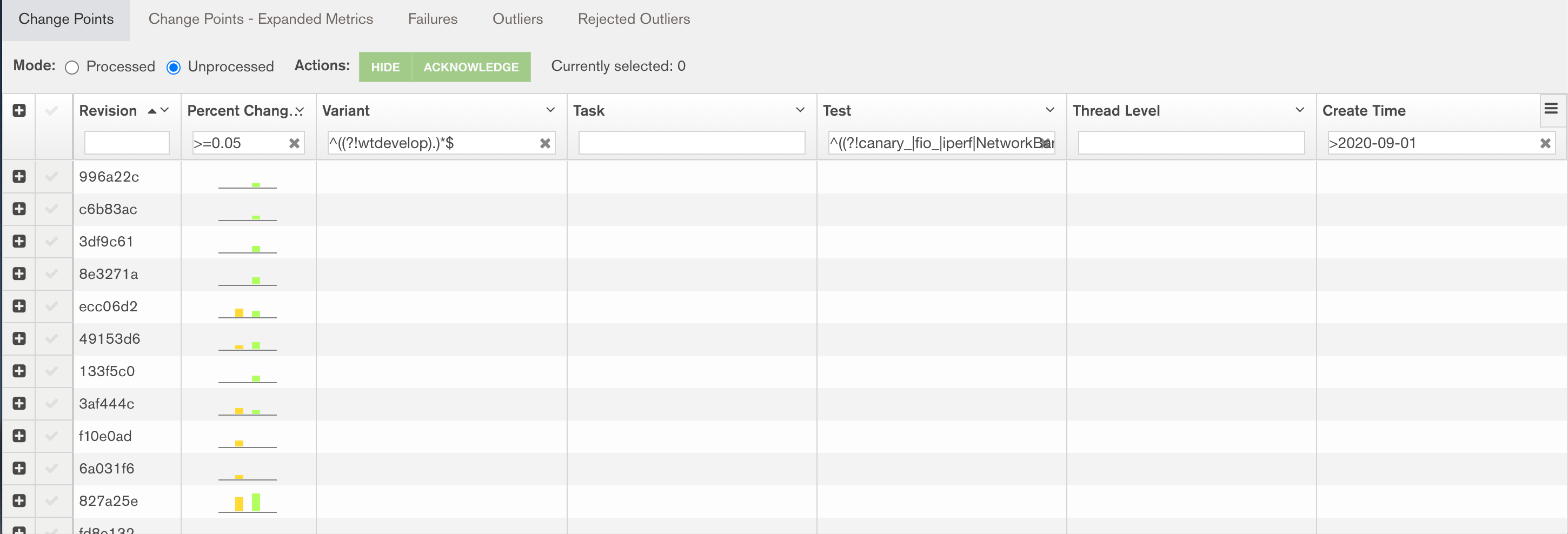}
  \caption{The existing build baron triage page is used by the build barons to
    triage change points on the existing data.}\label{fig:triage-existing}
  \Description{Screenshot of the existing build baron triage page, which is used by the build barons
  to triage change points on the existing data}
\end{figure*}

\begin{figure*}[tb]
  \centering
  \includegraphics[width=\linewidth]{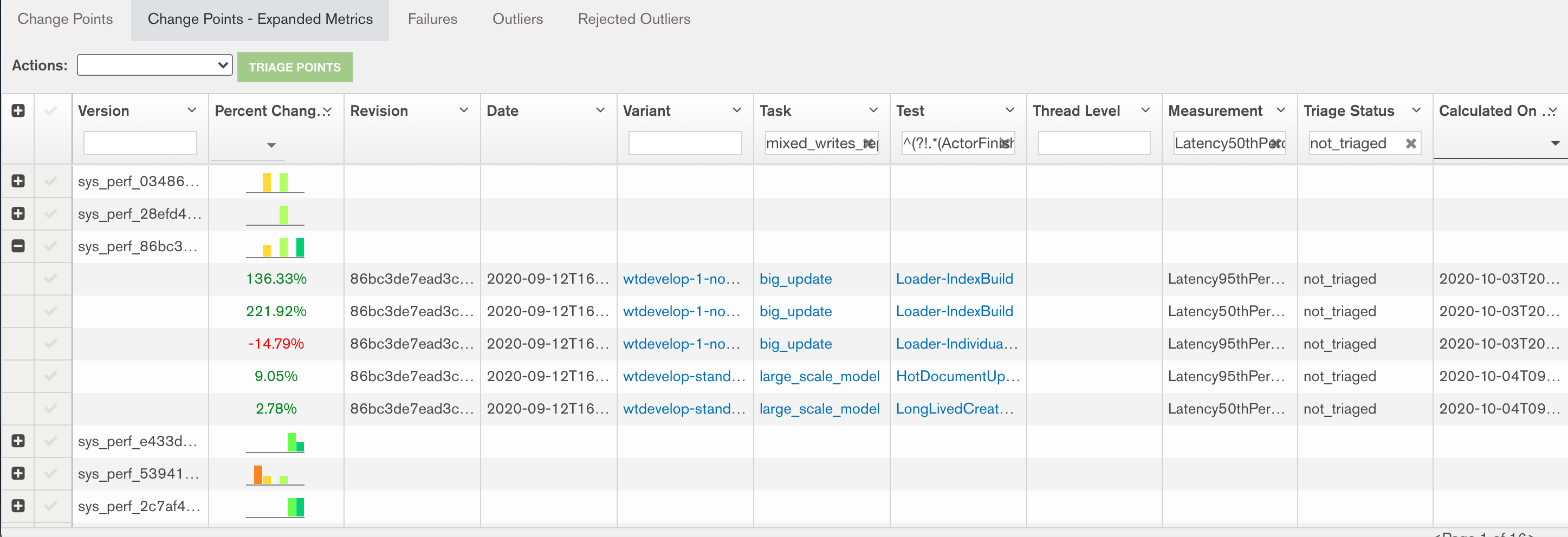}
  \caption{The new build baron triage page is used by the build barons to triage change points
    detected on the new, expanded metrics.}\label{fig:triage-expanded}
  \Description{Screenshot of the new build baron triage page, which is used by the build barons to
    triage change points for the new, expanded metrics}
\end{figure*}

Figure~\ref{fig:triage-existing} shows a snapshot of the existing build baron triage page and
Figure~\ref{fig:triage-expanded} shows a snapshot of the new triage board. These pages are setup to
enable the build barons to triage detected change points, create JIRA tickets, and assign those
tickets to teams. We aggregate all change points for a given commit revision into one line by
default to simplify processing. Each group of change points can be expanded to show all the impacted
tests and configurations, as is done for one group in Figure~\ref{fig:triage-expanded}.

For now we have placed all the new data on a new tab called ``Change Points - Expanded
Metrics''. Adding a new tab is not optimal, but it does allow us to update and experiment with the
new system with no fear of breaking our existing system and the processing of the legacy ``Change
Points'' tab. Eventually we expect that the two tabs will merge together. The new tab has the
additional column ``Measurement''. The argument in the field is a regular expression allowing tight
control and filtering for the build baron. For now, the system is setup to display three such
metrics ($50$th, $95$th, and $99$th percentile latencies). We expect to add more metrics to be
triaged, as well as migrating the legacy metrics to this page in the future. The page also shows for
each change the date the change was committed (Date) as well as the date on which the change point
was calculated (Calculated On). The first is useful for understanding the development of the
software, while the latter is useful for insight into the change point detection process. A change
point that has been calculated recently is the result of more recent test executions. Both dates
replace the somewhat ambiguous ``create time'' on the original page.

\begin{figure}
  \centering
  \includegraphics[width=\columnwidth]{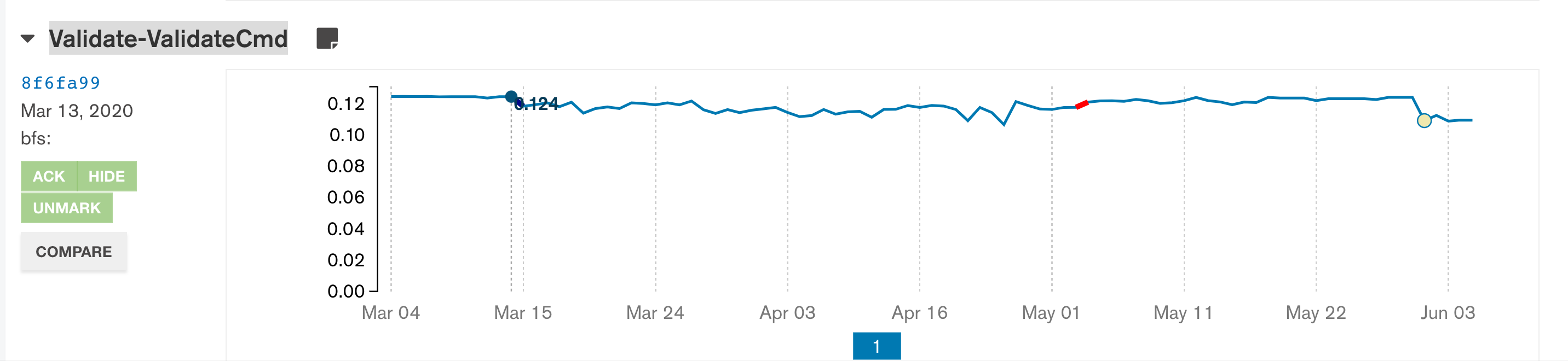}
  \includegraphics[width=\columnwidth]{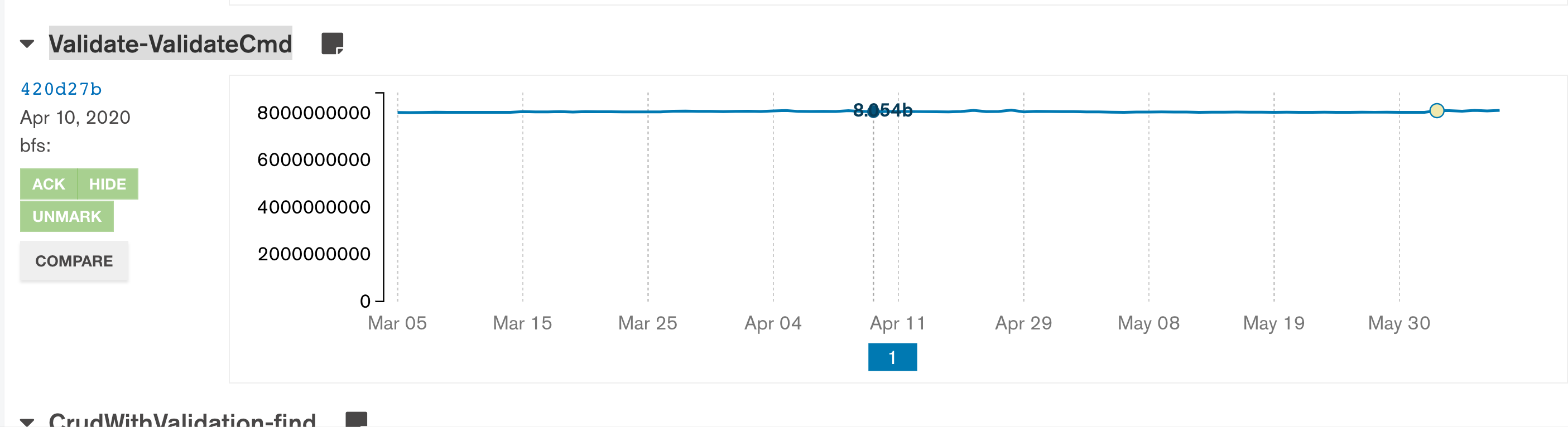}
  \includegraphics[width=\columnwidth]{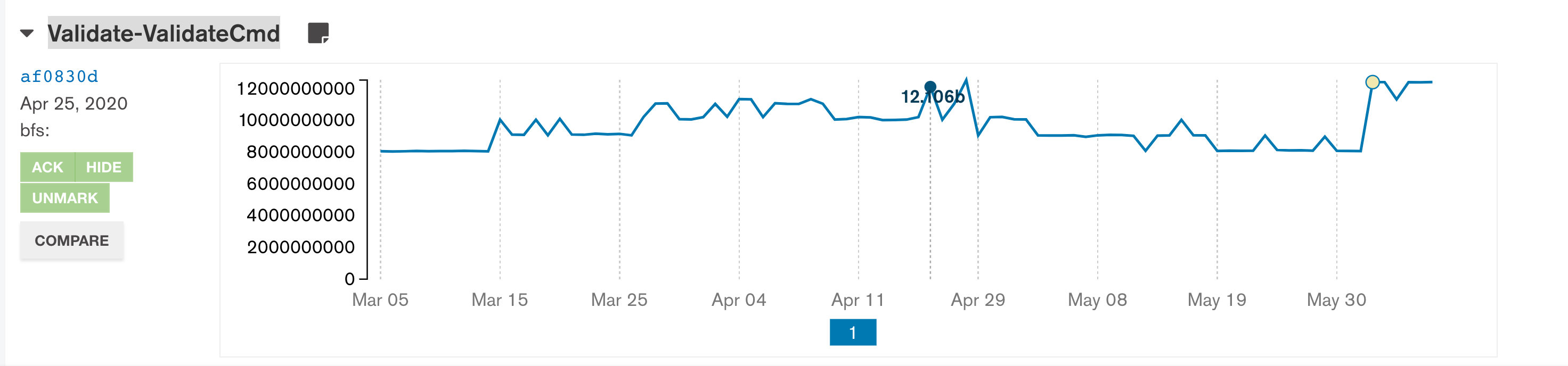}
  \caption{Three trend views of the same test, showing a performance regression. All three show
    performance over time. The top graph shows throughput, the middle shows median latency, and the
    bottom graph shows $90$th percentile
    latency. The regression is visible on the throughput and $90$th percentile latency graphs, but
    not for the median latency.}
  \Description{Graphs showing performance over time for one test. The top graphs shows a mostly flat
    line, followed by a small dip to a lower flat line. That graph shows the average throughput, so
    this is a performance regression. The second graph shows a completely flat line for the median
    latency for the test. The last graph shows a line with some variability, and an increase to a
    higher level towards the right side. That is 90\% latency, showing a clear increase, and is the
    cause of the drop in throughput.}
  \label{fig:trend-latency}
\end{figure}

We also display trend graphs for each test, showing the evolution for a performance result over
time, as the software is developed. The graphs are included on the page summarizing results for each
task. As in the case of the triage page, we worried about overwhelming the users with additional
results, so we added a pull down enabling the user to select which metric to
display. Figure~\ref{fig:trend-latency} shows a particularly interesting example of the value of
these additional metrics and graphs. We detected a small change in average throughput, but further
investigation showed a clearer change in the 90th percentile latency, while there was no change in
the median latency. This information makes it easier to debug the issue, as it clearly is not the
common path that is slower, but rather something making a small fraction of the operations
significantly slower.

\section{Lowering the Performance Impact of System Issues}\label{sec:systemissues}

We run many of our performance tests in the Cloud and have done work to reduce the noise and
increase the reproducibility of that system~\cite{henrik_ingo_reducing_2019}. Sometimes there are
performance problems on the testbed itself. We use ``canary tests'' to detect that. A ``canary
test'' is a test that tests the testbed instead of software under test. In normal operation we
expect the results for our canary tests not to change over time.

The canary tests are tests just like any other test, but treating them the same leads to some
challenges. First, anyone looking at the result needs to know what is a canary test and what is
not. We do not want server engineers spending any time diagnosing canary test failures. At the same
time, we also do not want a server engineer diagnosing a performance change on a non-canary test
when a canary test has also failed. Ideally, we would discard that result because it is suspect, and
rerun those performance tests. If we were able to completely discard every (or even most) case of
significant noise due to the system, it makes the job of the change point detection algorithm that
much easier.

We set out to lower the impact of system noise by leveraging the data from our change point
detection algorithm. We recognized that while changes in server performance manifested as changes in
the distribution on our performance test results, system noise was different. The common problem was
a bad run with results that did not match recent history. This is a problem of finding statistical
outliers, not of finding change points. As NIST defines it, ``An outlier is an observation that
appears to deviate markedly from other observations in the sample.''~\cite{noauthor_13517_nodate}.

There are a number of existing outlier detection algorithms. We implemented the Generalized ESD Test
(GESD)~\cite{rosner_percentage_1983, noauthor_135173_nodate} algorithm. The code is included in our
open source signal processing repository (https://github.com/10gen/signal-processing). Specifically,
we wanted to use the outlier detection to detect outliers on canary tests. An outlier on a canary
test would indicate something strange happened on the testbed. We want to not use the data from such
a run, and ideally rerun those experiments. When an outlier is detected on a canary test, we would
automatically suppress the test results for that task and reschedule the task.

\begin{figure}
  \centering
  \includegraphics[width=\columnwidth]{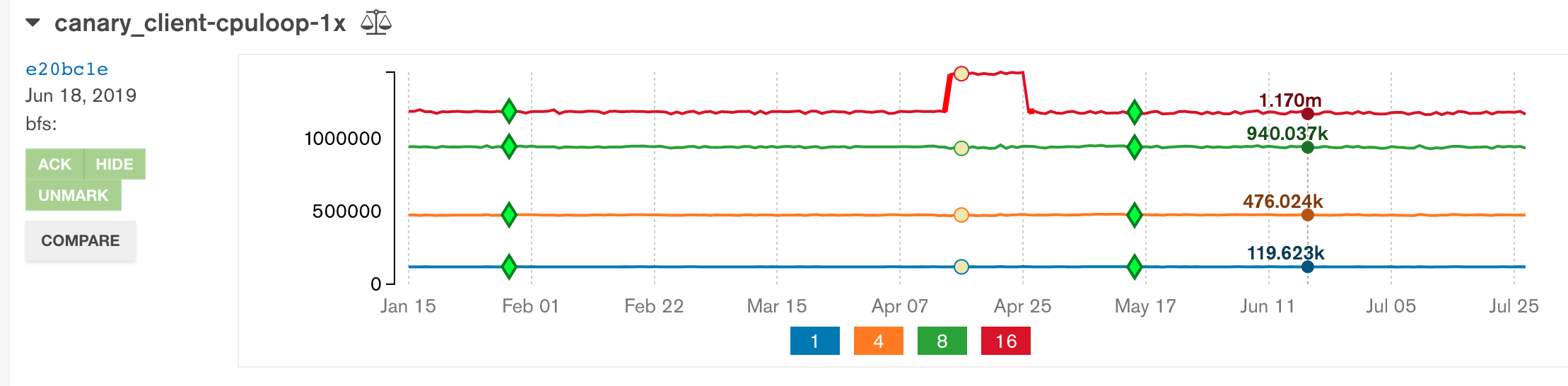}
  \caption{Performance for one of our canary workloads over time. It shows a real, if short lived,
    change in testbed performance, causing the outlier detection based system to rerun many tests.}
  \label{fig:canary-change}
\end{figure}

While reasonable in theory, we ran into some challenges. Figure~\ref{fig:canary-change} shows an
example of one such challenge: We had a short period of time in which the underlying system got
faster. This may have been a temporary configuration change. Essentially every task that ran after
that change was flagged as an outlier and re-run. In fact, they were all run 3 or more times. This
cost a lot of money and (worse) slowed our ability to get results. Also, as it was a real change in
the underlying testbed performance, the results did not noticeably change with any of the
re-runs. In this case we spent a lot of money for no improvement. We added a system to ``mute'' such
changes, but it required active intervention to avoid the worst cases. This change did not last
long, but it was long enough to cause more outliers and re-runs when the performance returned to
normal.

In other cases the system would rightly detect a transient change in testbed performance, but the
underlying issue lasted for some period of time. The tests would immediately rerun, but still get
the bad results. Only after waiting some period of time would the performance return to
normal on a rerun.

At the end of the day we disabled the system. It was not solving our problem, but it was costing us
money. We have kept the computation running, so we have built up a wealth of data when we come back
to this area or decide to use outlier detection for other challenges.

\section{Improved Comparison of Arbitrary Results}\label{sec:comparison}

Our previous work on change point detection~\cite{daly_use_2020} only addressed identifying when
performance changed. It did nothing for comparing two arbitrary builds to see if performance
changed. There are two common cases in which we want to compare arbitrary test runs:
\begin{enumerate}
\item Comparing performance from recent commits to the last stable release.
\item Comparing performance from a ``patch'' build (proposed change). Does that patch change
  performance?
\end{enumerate}

In the first case we want to determine the net performance change over a period of time. Very
commonly this is how we check how our proposed release compares to the previous release. We would
like to know what is faster, what is slower, and what is more or less stable now compared to
then. There may be multiple changes in performance for a given test across a release cycle. Change point detection
helps us understand each of those changes, but at the end of the day we need to let our customers
know what to expect if they switch to the newer version. This check also provides a backstop to change point detection
to make sure nothing significant has slipped through the triage process.

In the second case the engineer needs to know what impact their changes will have on performance. We
have tools to compare the results from two arbitrary test executions, but it does not have any sense
of the noise distribution for the test. It makes it hard to tell which differences are ``real'' and
which are just artifacts of the noise of those particular runs. A common pattern to deal with this
is to compare all the data, sort by percentage change, and inspect the tests with he largest
changes. Invariably the largest reported changes are due to noise, usually from tests that report a
low absolute result value (e.g., latency of something fast), leading to large percentage changes. An
advanced user may learn which tests to ignore over time, while a less experienced user may either
use brute force, or enlist an experienced user. Neither solution is a good use of time.

The change point detection system does not directly improve our ability to compare performance
across releases, however, its results do enable smarter comparisons. All of the data from the change
point detection algorithms is available in an internal database. That data includes the location of
change points, as well as sample mean and variances for periods between change points. The sample
mean averages out some of the noise, and the sample variance gives us a sense of how much noise
there is. We can use that data a number of ways to improve the comparison. The simplest may be to
compare means instead of points, and use the variance data to understand how big the change is
relative to regular noise.

After a few iterations we had the following system:
\begin{itemize}
\item Select two revisions to compare.
\item Query the database for all the raw results for each revisions.
\item For each result query the database for the most recent change point \textbf{before} the given
  revision. Save the sample mean and variance for the region \textbf{after} the change point.
\item Compute a number of new metrics based on those results.
\end{itemize}

The new computed values were:
\begin{itemize}
\item Ratio of the sample means
\item Percentage change of the sample means
\item Change in means in terms of standard deviation
\end{itemize}

Note that there are better statistical tests we could use (see future work in
Sec~\ref{sec:future}). Comparing means and standard deviations is not technically correct for
determining the probability that a change is statistically significant. However, it is both easy to
do and proved useful for a prototype.

We exported the data as a CSV file and operated on it in a spreadsheet for a first proof of
concept. Our first instinct was to sort all the results by how many standard deviations a change
represented, however, that did not work well. It turned out that some of our tests reported very low
variances. The top results ended up being very small changes in absolute terms, but huge changes in
terms standard deviation. With that in mind, we shifted to a more complex strategy: we filtered out
all results that were less than a 2 standard deviation change, and then sorted by percentage
change. We felt comfortable doing that since we did not need to catch every change for the current
use, only the most significant (in a business sense, not a statistical one) changes. A change that
was less than two standard deviations was unlikely to be the performance change that the engineering
organization had to know about. Once we filtered on number of standard deviations and sorted on
percentage change, the signal greatly improved. The most important changes rose to the top and were
reviewed first.

\begin{figure*}
  \centering
  \includegraphics[width=\linewidth]{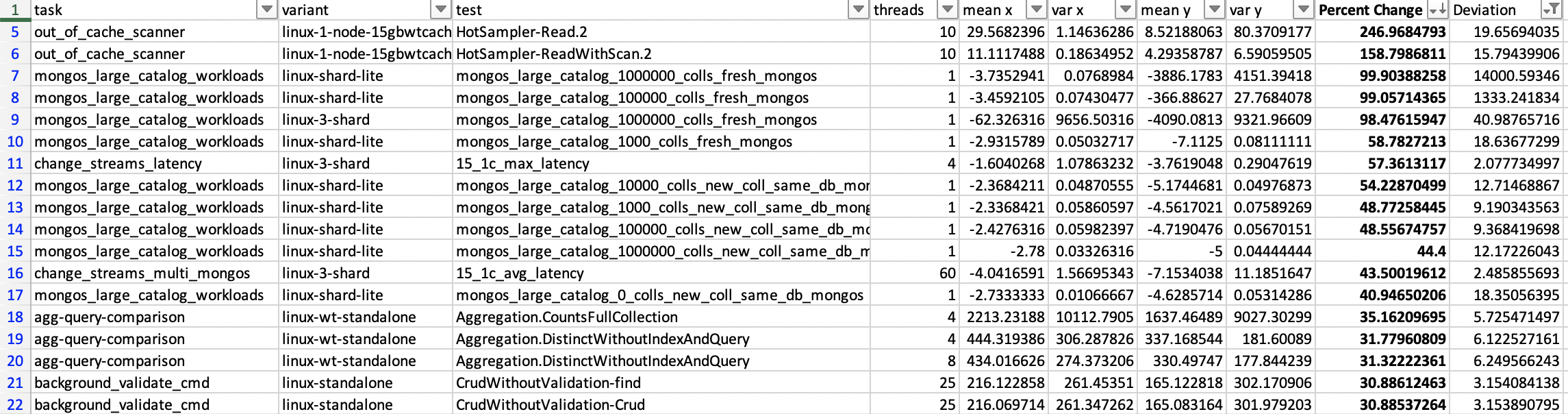}
  \caption{Spreadsheet view of performance comparing performance of two commits, taking advantage of
    the statistics generated by the change point detection algorithm.}\label{fig:perfdiscoverypoc}
\end{figure*}

We regularly need the ability to compare two commits as part of a monthly update on
performance. Once a month we checkpoint the current status of performance for the development branch
against the previous month, and against the last stable release. This gives us the big picture on
the state of performance, in addition to the detailed results from change point
detection. Figure~\ref{fig:perfdiscoverypoc} shows a spreadsheet we created using this process for a
recent monthly checkpoint on the state of performance. The figure shows the two standard deviation
filter applied (``Deviation'' column), and then sorted on the ``Percent Change'' column.  This view
enabled us to quickly review all the real changes and avoid changes that were due to noisy
tests. For example, the top test is $250\%$ faster across the comparison. While we have shown
performance improvements in the figure, we review both improvements and regressions to get a
complete view of the state of performance.

In practical terms, this POC has lowered the cost of reviewing the monthly build from multiple
hours, to somewhere between 30 and 60 minutes. Additionally, all of that time is now productive time
looking at real issues. If there are more issues, it takes more time, and if there are fewer, it
takes less time. We expect to transition this view from the CSV and proof of concept stage, into
another page in our production system available to all engineers. We also expect to implement more
rigorous statistical tests.

\section{Impact}\label{sec:impact}

The combination of the changes described above has had noticeable impact on our performance testing
infrastructure and on our engineering organization. The basic way we track a performance change is a
JIRA tickets. We compiled statistics from our JIRA tickets to quantify part of that impact. The
statistics are aligned with our release cycle, which is nominally a year long.

\begin{table}
  \centering
  \begin{tabular}{lrr}
    \toprule
    \textbf{Release Cycle} & \textbf{4.2.0} & \textbf{4.4.0} \\
    \midrule
    Total Tickets & 273 & 393 \\
    Resolved Tickets & 252 & 346 \\
    Percent Resolved & 92.3\% & 88.0\% \\
    Resolved by Release & 205 & 330 \\
    Percent Resolved by Release & 75.1\% & 84.0\% \\
    Release Duration & 412 & 352 \\
    Tickets per Day & 0.66 & 1.12 \\
    \bottomrule
  \end{tabular}
  \caption{Statistics on performance related JIRA tickets over the previous two release cycles.}
  \label{tab:tickets}
\end{table}
We had considerably more performance related BF tickets in 4.4.0 than 4.2.0, over a shorter release
cycle. Tickets per day went from $0.66$ to $1.12$, a $70\%$ increase. We had a large increase in
tickets, but simultaneously increased the percentage of tickets resolved by the release. Those are
both positive signs, especially since we spent the same amount of time triaging those changes, but
it is only truly positive if the ticket quality has stayed the same or improved.

\begin{table}
  \centering
  \begin{tabular}{lrr}
    \toprule
    \textbf{Release Cycle} & \textbf{4.2.0} & \textbf{4.4.0} \\
    \midrule
    Code related & 28.57\% & 43.06\% \\
    Test related & 8.73\% & 7.80\% \\
    Configuration related & 0.00\% & 0.58\% \\
    System related & 28.17\% & 24.86\% \\
    Noise related & 7.94\% & 6.94\% \\
    Duplicate ticket & 11.11\% & 14.45\% \\
    Not labeled & 16.67\% & 2.31\% \\
    \bottomrule
  \end{tabular}
  \caption{Breakdown of root causes for performance JIRA tickets.}
  \label{tab:tickquality}
\end{table}

Table~\ref{tab:tickquality} shows quality information about our performance tickets. We label every
ticket based on its cause. The best case is for the change to be code related: that indicates that
the ticket captures a performance change based on changes in the code under test. These are tickets
telling us something useful about our software. There are many other causes for tickets
however. Performance changes can be created due to changes in the test (test related) or the testbed
configuration (configuration related), the system itself can directly cause an error (system
related), or noise in the system can create a false alert (noise related). Sometimes we create
multiple tickets which we eventually determine are the same cause (duplicate ticket). Finally, some
tickets are not labeled at all because they do not have a clear cause.

The fraction of code related tickets has gone up, even as the ticket volume has also gone up. We can
conclude that we are generating more tickets, with the same amount of time dedicated to triage, and
the tickets are of higher quality than last year. In other words, we are doing our job better than
last year. While we are happy with that improvement, we also recognize that less than half our
tickets are about changes in the software under test. We would like to continue to drive that
percentage higher.

Interestingly, the category with the largest drop are tickets that are not labeled. This is due to
us doing a better job of diagnosing tickets and making them actionable. It is not the case that we
were just missing code related tickets with the labels in the past. The number of duplicates is the
only non-code related category to go up noticeably. We attribute this to the increase load of change
points and tickets on the build barons.

\begin{table}
  \centering
  \begin{tabular}{lrr}
    \toprule
    \textbf{Release Cycle} & \textbf{4.2.0} & \textbf{4.4.0} \\
    \midrule
    Performance improvements & 21 & 40 \\
    Percentage of tickets that are improvements & 7.69\% & 10.18\% \\
    Days per performance improvement & 19.62 & 8.80 \\
    Performance regressions & 15 & 13 \\
    Percentage of tickets that are regressions & 5.49\% & 3.31\% \\
    Days per performance regression & 27.47 & 27.08 \\
    \bottomrule
  \end{tabular}
  \caption{Breakdown on the number and rate of performance JIRA tickets closed as improvements and
    regressions over the past two release cycles.}
  \label{tab:tickimprovement}
\end{table}

The last measure of goodness is how many tickets were fixed (or not), and how many things
improved. Table~\ref{tab:tickimprovement} shows those statistics. Before discussing the numbers we
note that we count any net improvement as an improvement and any net regression closed without
fixing as a regression, regardless of its practical significance. We had comparable number of
accepted regressions year over year, while nearly doubling the number of improvements. So, even with
the large increase in tickets, we still only get a regression that is not fixed about once a month,
and we went from getting an improvement every 20 days to one every 9 days.

Clearly our system is working better. We have more tickets and they are higher quality. In addition
to each component performing better, we believe that we have enabled a virtuous cycle. Performance
issues get diagnosed faster, making them easier to fix, so more issues get fixed. Development
engineers get used to receiving performance tickets and know they are high quality and
operational. Since the system provides useful information, engineers are more likely to look to fix
their regressions and to add more performance tests. As we add more performance tests, we are more
likely to catch performance changes. One last improvement is that with increased trust, engineers
are more likely to test their key changes before merging, so we can avoid some performance
regressions ever being committed to the development mainline.

\section{Future Work and Continuing Challenges}\label{sec:future}

Our current performance testing system enables us to detect performance changes during the
development cycle, and to enable our developers to understand the impact of their code
changes. While we have made great progress, there is still much that we would like to improve in the
system. We expect that everything (commits, tests, results, changes) will continue to increase,
putting more load on our system. Additionally, we are increasing our release frequency to
quarterly~\cite{mat_keep_accelerating_2020}, which will further increase the load on the system. In
the near term we are working to improve the ability to compare arbitrary versions, building on the
work described in Section~\ref{sec:comparison}. This will involve both using better statistical
tests, such as Welch's t-test~\cite{welch_generalization_1947} (assuming normality) or Mann-Whitney
U-test~\cite{mann_test_1947} in place of the simple variance based calculation, as well as building
the view into our production system. This will help us to compare performance between releases, as
well as help developers determine if their proposed changes impact performance.

There is still much we can do on the change point detection itself. In order to simplify the
implementation, all tests and configurations are treated as separate and independent time series by
the change point detection algorithm. We think there is a large opportunity to consider correlations
between tests and configurations. It is very infrequent that one test and configuration changes
separately from all others. We should be able to exploit correlated changes to better diagnose and
describe real performance changes, and exclude noise.

There is still too much noise in the system, including some cases of particularly nasty noise. Two
examples include tests that show bimodal behavior and unrelated system noise. Some tests will return
one of two different results, and may stay with one of those results for a period of time before
reverting to the other (e.g., 5 tests runs at $20$ followed by 4 tests runs at $10$). The change
point detection algorithm has a very hard time with bimodal behavior as it looks like a statistical
change. Today, a human has to filter these changes out. There are also cases of system noise that
are real performance changes due to compiler changes. Sometimes these are due to code layout issues
letting a critical code segment fit within or not fit within a performance-critical hardware
cache. These issues manifest as deterministic changes in performance, but there is not much we can
do about them except filter them out by hand.

Ultimately, the goal of all of this work can be described as a multi-dimensional optimization problem. We want to
simultaneously:
\begin{itemize}
\item Maximize the useful signal on performance versus noise and distractions.
\item Maximize the test and configuration coverage.
\item Minimize the cost of performance testing.
\item Minimize the time from creation of a performance change to its detection, diagnosis, and
  fix. (the limit of this is catching a regression before commit).
\end{itemize}
We have work to do on all of these points. Often, in the past, we have found ourselves with bad
options, which explicitly trade off one point for another. We hope to develop techniques that
improve one or more items above at the same time, without hurting the others.

\section{Related Work}
\label{sec:related}
Related work has looked at testing performance in continuous integration systems. Rehman et
al.~\cite{rehmann_performance_2016} describe the system developed for testing SAP HANA and stressed
the need for complete automation. The system compared results to a user specified limit in order to
determine a pass fail criterion. The authors also discuss challenges in reproducibility, isolation,
and getting developers to accept responsibility for issues.

Continuous integration tests need to be fast, but standard benchmarks require extended periods of
time to run. Laaber and Leitner~\cite{laaber_evaluation_2018} looked at using microbenchmarks for
performance testing in continuous integration to deal with this problem. They found some, but not
all microbenchmarks are suitable for this purpose.

Once performance tests are included in a CI system, the next challenge is to efficiently isolate the
changes. Muhlbauer et al.~\cite{muhlbauer_accurate_2019} describe sampling performance histories to
build a Gaussian Process model of those histories. The system decides which versions should be
tested in order to efficiently build up an accurate model of performance over time and to isolate
abrupt performance changes. The paper addresses a problem similar to our previous work on detecting
change points in test histories~\cite{daly_use_2020}, although our previous work assumes performance
test results have a constant mean value between change points.

Test result noise is an ongoing challenge. Several papers investigate both sources of
noise~\cite{duplyakin_datacenter_2020, maricq_taming_2018} and quantifying the impact of that
noise~\cite{laaber_software_2019}. Duplyakin et al.~\cite{duplyakin_datacenter_2020} use change
point detection to identify when the performance of the nodes in a datacenter change. Their
objective is to identify and isolate those performance changes in order to keep them from impacting
experiments run in the datacenter. The paper by Maricq et al.~\cite{maricq_taming_2018} includes a
number of practical suggestions to reduce performance variability. The suggestions should be useful
for anyone running performance benchmarks, and we perform many of these suggestions in our
system. They also show the lack of statistical normality in their results, validating our design
choice to not assume normality. Finally, Laaber et al.~\cite{laaber_software_2019} compare the
variability of different microbenchmark tests across different clouds and instance types on those
clouds, demonstrating that different tests and different instance types have wildly different
performance variability. Running benchmark test and control experiments on the same hardware can
help control the impact of that noise.

The related area of energy consumption testing shows similar issues with test noise. Ournani et
al.~\cite{ournani_taming_2020} describe the impact of CPU features (C-states, TurboBoost, core
pinning) on energy variability. We have observed similar impacts on performance variability from
those factor in our test environment~\cite{henrik_ingo_reducing_2019}. Other work looks at extending
the state of the art for change point detection in the presence of
outliers~\cite{paul_fearnhead_changepoint_2019}. Our system is sensitive to outliers in the results
as well. Our efforts on outlier detection would have helped reduce the impact of outliers in our use
case, if it had been successful.

Finally, there is ongoing work related to our ultimate goal of more efficiently detecting changes
while simultaneously increasing our overall performance test coverage. Grano et
al.~\cite{grano_testing_2019} investigated testing with fewer resources. While this work is focused
on correctness testing, the principles can be extended to performance testing. Multiple
papers~\cite{de_oliveira_perphecy_2017, huang_performance_2014} try to identify which software
changes are most likely to have performance impact in order to prioritize the testing of those
changes. Huang et al.~\cite{huang_performance_2014} use code analysis of software changes to decide
which changes are most likely to impact which tests, while de Oliveria et
al.~\cite{de_oliveira_perphecy_2017} use many indicators (including static and dynamic data) to
build a predictor of the likelihood of a performance change in the tests based on a given software
change. Other work has focused on efficiently finding performance changes across both versions and
configurations~\cite{muhlbauer_identifying_2020} and is specifically focused on minimizing test
effort while enabling the testing of potentially huge space of configuration options and software
changes. We hope to build on these efforts to improve the efficiency of our performance testing.

\section{Acknowledgments}
\label{sec:acknowledgements}

The work described in this paper was done by a large collection of people within MongoDB\@. Key
teams include the Decision Automation Group (including David Bradford, Alexander Costas, and Jim
O'Leary) who are collectively responsible for all of our analysis code, the Server Tooling and
Methods team who own the testing infrastructure, the Evergreen team which built Cedar and Poplar for
the expanded metrics support, and of course our dedicated build baron team whom make the whole
system work.

We would also like to thank Eoin Brazil for his feedback on drafts of this paper.

\section{Conclusion}
\label{sec:conclusion}

In this paper we have reviewed a number of recent changes to our performance testing infrastructure
at MongoDB\@. This builds on previous work we have done to automate our performance testing
environment, reduce the noise in the environment (both actual noise and its impact), and better
makes use of the results from our performance testing. This infrastructure is critical to our
software development processes in order to ensure the overall quality of the software we develop.

We first reviewed the general increase in load on the infrastructure. Each year we run more tests in
more configurations while our developers commit more changes to our source repository. Overall we
had a more than $3x$ increase over two years in the total possible number of test results to
generate and analyze.

Paired with the general increase in load, we focused on improving the scalability of our ability to
process those results and isolate performance changes. We also added the ability to report more and
more descriptive results from tests, enabling saving information about every operation within a
performance test. This required new systems to store and process the results, as well as new
displays for triaging the results.

Attempting to better control system noise, we built a system to detect when the performance of our
testbeds changed, and therefore we should not trust the results of our performance tests. While
promising in theory, in practice this did not work as well as we had hoped, and ultimately we
disabled it.

Finally, we enabled better comparison of results between arbitrary commits. This was a large open
challenge for us. Building upon the change point detection system we use to process our results, we
were able to give a much clearer view of the significant changes between arbitrary commits, making
it much easier to regularly check the current state of the development software against the last
stable release. We continue to both refine this comparison of results and lift it into our
production environment.

The cumulative impact of these changes and our previous work has been to enable a virtuous cycle for
performance at MongoDB\@. As the system is used more, we catch and address more performance changes,
leading us to use the system more. This virtuous cycle directly increases the productivity of our
development engineers and leads to a more performant product.

 \bibliographystyle{ACM-Reference-Format}
 \bibliography{virtuous}
\end{document}